\documentclass[sigconf]{acmart}

\AtBeginDocument{%
  \providecommand\BibTeX{{%
    \normalfont B\kern-0.5em{\scshape i\kern-0.25em b}\kern-0.8em\TeX}}}

\usepackage[utf8]{inputenc} % allow utf-8 input
\usepackage[T1]{fontenc}    % use 8-bit T1 fonts
\usepackage{hyperref}       % hyperlinks
\usepackage{url}            % simple URL typesetting
\usepackage{booktabs}       % professional-quality tables
\usepackage{amsfonts}       % blackboard math symbols
\usepackage{nicefrac}       % compact symbols for 1/2, etc.
\usepackage{microtype}      % microtypography
\usepackage{xcolor}         % colors
\usepackage{caption}
\usepackage{subcaption}
\usepackage{enumitem}
\usepackage{natbib} % has a nice set of citation styles and commands
\usepackage{mathtools} % amsmath with fixes and additions
\usepackage{booktabs} % commands to create good-looking tables
\usepackage{tikz} % nice language for creating drawings and diagrams
\usepackage{amsthm}

\usepackage{comment}
\usepackage{url}
\usepackage{kantlipsum}
\usepackage{enumitem}
\allowdisplaybreaks

\newcommand{\indep}{\perp \!\!\! \perp}

\newtheorem{remark}{Remark}

\usepackage{graphicx,subcaption}
\newcommand\newsubcap[1]{\phantomcaption%
       \caption*{\figurename~\thefigure(\thesubfigure): #1}}

\copyrightyear{2023} 
\acmYear{2023} 
\setcopyright{acmlicensed}\acmConference[FAccT '23]{2023 ACM Conference on Fairness, Accountability, and Transparency}{June 12--15, 2023}{Chicago, IL, USA}
\acmBooktitle{2023 ACM Conference on Fairness, Accountability, and Transparency (FAccT '23), June 12--15, 2023, Chicago, IL, USA}
\acmPrice{15.00}
\acmDOI{10.1145/3593013.3594075}
\acmISBN{979-8-4007-0192-4/23/06}

\begin{document}

\title{Disentangling and Operationalizing AI Fairness at LinkedIn}


\author{Joaquin Quiñonero-Candela}
\authornote{Corresponding author.}
\email{joaquin@linkedin.com}
\affiliation{%
  \institution{LinkedIn}
  \country{USA}
}

\author{Yuwen Wu}
\email{yuwwu@linkedin.com}
\affiliation{%
  \institution{LinkedIn}
  \country{USA}
}
\author{Brian Hsu}
\email{bhsu@linkedin.com}
\affiliation{%
  \institution{LinkedIn}
  \country{USA}
}
\author{Sakshi Jain}
\email{sjain2@linkedin.com}
\affiliation{%
  \institution{LinkedIn}
  \country{USA}
}
\author{Jen Ramos}
\email{jeramos@linkedin.com}
\affiliation{%
  \institution{LinkedIn}
  \country{USA}
}
\author{Jon Adams}
\email{jnadams@linkedin.com}
\affiliation{%
  \institution{LinkedIn}
  \country{USA}
}
\author{Robert Hallman}
\email{rhallman@linkedin.com}
\affiliation{%
  \institution{LinkedIn}
  \country{USA}
}
\author{Kinjal Basu}
\authornote{Work done while the author was with LinkedIn.}
\email{basukinjal@gmail.com}
\affiliation{%
  \institution{LinkedIn}
  \country{USA}
}

\renewcommand{\shortauthors}{Quiñonero-Candela, et al.}

\begin{abstract}
Operationalizing AI fairness at LinkedIn's scale is challenging not only because there are multiple mutually incompatible definitions of fairness but also because determining what is fair depends on the specifics and context of the product where AI is deployed. Moreover, AI practitioners need clarity on what fairness expectations need to be addressed at the AI level. In this paper, we present the evolving AI fairness framework used at LinkedIn to address these three challenges. The framework disentangles AI fairness by separating out equal treatment and equitable product expectations. Rather than imposing a trade-off between these two commonly opposing interpretations of fairness, the framework provides clear guidelines for operationalizing equal AI treatment complemented with a product equity strategy. This paper focuses on the equal AI treatment component of LinkedIn’s AI fairness framework, shares the principles that support it, and illustrates their application through a case study. We hope this paper will encourage other big tech companies to join us in sharing their approach to operationalizing AI fairness at scale, so that together we can keep advancing this constantly evolving field.
\end{abstract}




\maketitle

\section{Introduction}
\label{sec:intro}

LinkedIn’s products aim to live up to the company’s vision to \textit{“create economic opportunity for every member of the global workforce”}\footnote{https://about.linkedin.com/}. This would be impossible without leveraging AI at scale. We use AI to power many product offerings, whether to recommend new job openings to job seekers \cite{kenthapadi2017personalized}, show qualified candidates to recruiters \cite{geyik2018talent, ramanath2018towards}, share relevant news and articles to our members \cite{agarwal2014activity, agarwal2015personalizing, agarwal2018online}, or recommend new connections to help members grow their network \cite{guy2016people}. Leveraging AI for a plethora of products and creating opportunities for every member means it is critical to work towards building AI that treats every member fairly.

But what does it mean for AI to be fair, or not biased? Who should decide what definition of AI fairness applies to a given product? Given a definition of AI fairness, how do we operationalize measurement and mitigation at scale? These questions aren’t rhetorical, they are motivated by substantial challenges we have found when trying to operationalize AI fairness at LinkedIn's scale. These challenges include:

\begin{itemize}%[noitemsep,topsep=0pt,parsep=0pt,partopsep=0pt]
    \item The multiple reasonable definitions of AI fairness that are mutually incompatible, and choosing one vs. another is consequential and dependent on the product \cite{corbett2018measure,narayanan21fairness,raghavan2020mitigating}.
    \item Choosing a particular definition of AI fairness for a specific product requires a deep understanding of the context in which that product is deployed, an assessment of the associated benefits, risks and potential unintended consequences, and thoughtful oversight and governance \cite{crawford2021atlas}.
\end{itemize}

Imagine a recruiter searches for qualified candidates for a job opening where gender is irrelevant to being qualified for this job. The recruiter gets a list of 100 candidates with 80 males and 20 females (we’re using binary gender in this example). Is this unfair because we didn’t get 50 males and 50 females, or something closer to a 50/50 distribution? But would 50/50 be fair if in the current social context, of all people who are qualified for this job (e.g. have the relevant skills) only 20\% are female? Should an AI powering recruiter search aim to accurately predict whether a candidate is qualified, irrespective of their gender, reflecting potential societal imbalances?\footnote{Note that in this thought example, the AI is an assistive tool for the recruiter, it doesn’t make hiring decisions, the recruiter does.} Or should the AI instead take gender into consideration and compensate for societal imbalances by increasing female representation?

One way to reason about this tension is to consider the difference between equality and equity when we think about what is fair. In our example, a principle of equality would mean treating every candidate the same, and ensuring the AI predictions correctly predict real-world qualifications as accurately for female as for male candidates. In contrast, a principle of equity would investigate whether there have been historical barriers to females acquiring the qualifications required for the job, or whether the qualifications required are too narrow and would exclude females capable of performing that job as well as males. An equity principle may require having a female representation in the search results that is higher than the 20\% baseline. It is important to note that a female representation higher than 20\% would require treating female candidates differently than male ones explicitly based on their gender, and therefore would be incompatible with a principle of equality (or equal treatment).

The tension between equality and equity is mirrored in the inherent tradeoffs between AI fairness metrics like predictive parity (or calibration) which aligns with equal treatment, and demographic parity (and to some extent equalized odds) which align with equitable outcomes, see \cite{kleinbergFairnessTradeoff}. AI practitioners working on AI fairness are faced with a very difficult choice: what definition of AI fairness to use for measurement and mitigation? This decision boils down to making a consequential tradeoff on the spectrum between equal treatment and equitable outcomes. We take the position that:
\begin{enumerate}[noitemsep,topsep=0pt,parsep=0pt,partopsep=0pt]
    \item AI practitioners cannot be expected to have all necessary understanding of the context in which a given product is deployed, and therefore should not be put in a position to make consequential AI fairness tradeoff decisions.
    \item Addressing AI fairness comprehensively requires acting both on the AI components and also on the broader product powered by the AI, including product goals, policies and design as part of a comprehensive product equity strategy.
\end{enumerate}
We therefore propose to disentangle AI fairness by clearly separating out equal treatment and equitable outcome expectations, so that both can be addressed explicitly, and so AI practitioners have clear guidelines for operationalizing AI fairness at scale. Our framework can be summarized with this simplified equation:
\begin{align*}
\textrm{\textbf{AI Fairness}} = \textrm{\textbf{Equal AI Treatment}} + \textrm{\textbf{Product Equity}}.    
\end{align*}
This approach demands of the AI component that it treats everyone equally (we will detail what we mean by equal AI treatment in the remainder of the paper), but it also acknowledges that equal treatment is not sufficient to ensure equitable product experiences and outcomes. Hence, our AI Fairness approach requires a complementary product equity strategy that considers additional efforts focused on understanding potential harms or barriers that disproportionately impact certain groups, as well as potential investments that can make the product accessible and beneficial to all.

In the example above, an AI that satisfies equal AI treatment will not address the social problem of female under-representation in the pool of qualified candidates, and as a result, females will still be underrepresented in the search results. A complementary product equity strategy will for example invest in awareness and actionable suggestions for recruiters to increase the diversity of their candidate pool, such as “diversity nudges” that let recruiters know females are underrepresented in their search, and suggestions to consider additional relevant skills, and broader or different search criteria \cite{diversityNudges}. Another complementary product equity strategy would be to re-rank the candidates so every slate of candidates (say the 20 candidates that fit in each of the 5 pages that together contain all candidates) matches the proportions of the underlying qualified candidate distribution \cite{geyik2018talent}.
 
In this paper, we focus on the equal AI treatment component of our AI Fairness framework and share the principles that support it. While we emphasize where a product equity strategy needs to complement equal AI treatment, we aim to present details of our product equity strategy in future work. Meta and Google have shared aspects of their approaches to AI fairness in \citet{bakalar2021fairness} and \citet{beutel2019putting} respectively. Like our work, Meta's approach offers a comprehensive end to end framework for thinking about AI fairness. Although it also builds on a notion of predictive parity, it focuses on a binary classification fairness criterion around the decision boundary. Google's approach clearly articulates a measurement and mitigation approach, but it does not capture the full process of evaluating and selecting the appropriate mitigation strategy, or potential complementary product changes. Like Meta, it focuses on the binary classification case, but with a different definition of fairness, "conditional equality," that extends equality of opportunity and therefore aims to close false positive rate (FPR) gaps. What distinguishes our work is that unlike in Meta or Google's case, we offer a justifiability framework for AI unfairness mitigation and demonstrate its application on a real-world LinkedIn product in Section \ref{sec:casestudy}.

We recognize that AI fairness is an evolving and complex area, with no consensus on definitions, goals, or mitigation strategies. We expect that our principles and our implementation strategies will develop over time, and we hope that an increasing number of large public companies will also share their AI fairness strategy. The rest of the paper is organized as follows. We first present LinkedIn’s principles for equal AI treatment in Section \ref{sec:principles}. Section \ref{sec:practice} then details the operationalization guidelines that allow the applications of these principles in practice. We illustrate the principles and their application through a real-world case study and share learnings and results in Section \ref{sec:casestudy}. We end with a discussion in Section \ref{sec:discussion}.

\section{Equal AI Treatment Principles at LinkedIn}
\label{sec:principles}
When developing principles for equal AI treatment at LinkedIn, we started by disentangling AI fairness by separating equal treatment and equitable outcome expectations. We set equal treatment expectations as the default bar for the AI component of our products so AI practitioners have clear guidelines that can be operationalized at scale. We then addressed the important challenge that equal AI treatment is not sufficient for achieving equitable outcomes and made sure the principles reflect the need for a complementary product equity strategy. Finally, recognizing that AI fairness is a fast evolving field, we committed to evolving our AI fairness strategy transparently and with the input of external stakeholders. These are LinkedIn’s Equal AI Treatment principles:

\begin{enumerate}
\item \textbf{We will measure and work to mitigate algorithmic bias so that our AI systems treat everyone equally.} This means measuring and, where appropriate, mitigating systematically unequal predictions or errors affecting member demographics. In other words, similarly qualified members should receive similar opportunities irrespective of what group they belong to. This principle does not extend to interventions intended to create equitable product experiences or outcomes.

\item \textbf{We will not consider equal AI treatment the end of our work but will treat it as the foundation of a broader fairness and equity strategy.} Additional measures – such as product features and design changes – and coordination between product and AI teams are key to achieving LinkedIn’s vision of  “creating economic opportunity for every member of the global workforce”, and to addressing unintended consequences of our equal AI treatment efforts.

\item \textbf{We will validate our approach externally and lead with transparency in this developing field.} Across our equal AI treatment and our broader equity and fairness initiatives, we will share learnings and case studies, and we will leverage collaboration and feedback from members, customers, advocacy groups, and social scientists. We aspire to learn from and guide other organizations in delivering equal AI treatment and exploring broader equity initiatives. 
\end{enumerate}

\section{From Principles to Practice}
\label{sec:practice}

The three principles above are the foundation on which we develop practical strategies to make sure the AI in our products treats every member fairly. In this section, we share the operationalization guidelines for the principles as a template that’s applicable across all AI-powered products at LinkedIn.

%The above three principles give us the foundation to develop the strategies towards realizing our overall goal of making all LinkedIn products fair to our members. In this section, we deep-dive into the implementation details of how each of the principles is being brought into practice across the entire company.

\subsection{Principle 1: We will measure and work to mitigate algorithmic bias so that our AI systems treat everyone equally}

\subsubsection{Definitions and Measurement:} 
\begin{figure}[!t]
    \centering
    \includegraphics[width=0.48\linewidth]{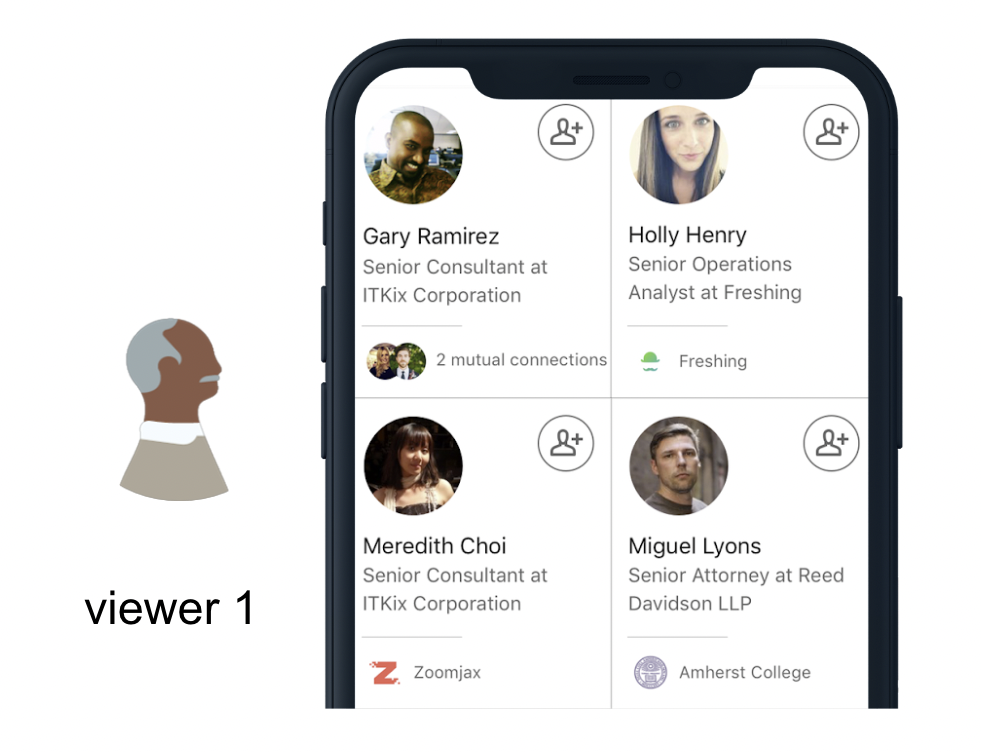}
    \includegraphics[width=0.45\linewidth]{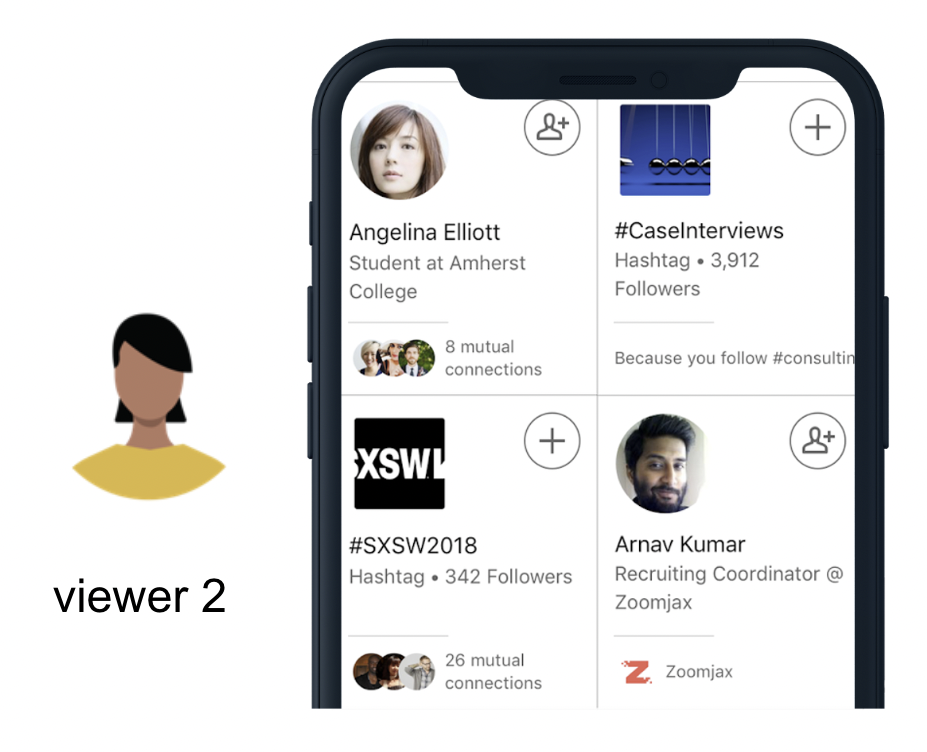}
    \caption{Viewers are members who get to see the recommendations. Here viewer 1 (on left) and viewer 2 (on right) see different recommendations. Equal AI treatment for them is through the standard quality of service. Recommended members are those that are being shown to viewers. For these recommended members equal AI treatment means equal opportunity for equally qualified and is measured by predictive parity.}
    \label{fig:two-sided}

\end{figure}

Equal AI treatment means that AI predictions treat people the same way, from a statistical perspective, irrespective of their demographic group membership. Conversely, algorithmic or AI bias means that people are not being treated the same way across groups. This could mean that the accuracy of predictions is systematically worse for a particular demographic group, or that systematic errors benefit some groups and disadvantage others. 
%Choosing to define algorithmic fairness as “equal AI treatment” informs our definition of algorithmic bias.
Formally, we define and measure equal AI treatment in different AI systems slightly differently depending on the context, but it relies on appropriate and accurate demographic group data in all cases. 
\begin{itemize}
\item When AI allocates scarce opportunities, equal AI treatment by default means “equal opportunity for equally qualified candidates”. Mathematically, we define it as predictive parity, where equally qualified candidates get similar predicted scores irrespective of what group they belong to. Formally, this is expressed as, 
$Y \indep A | \hat{Y}$ where, $\hat{Y}$ is the prediction score, $A$ is the random variable signifying the demographic group membership and $Y$ is the outcome of interest. For details see \cite{diciccio2022predictive, barocas2017fairness}. Product examples here include members who are ranked in People You May Know (PYMK) \cite{pymkblog}, Recruiter Search \cite{geyik2018talent}, as well as content being ranked in the Newsfeed \cite{agarwal2015personalizing}, where we are considering fairness with respect to content creators. We also differentiate how predictive parity is computed depending on whether the AI model does pointwise \cite{diciccio2022predictive} or listwise inference \cite{roth2022outcome}. 

\begin{remark}
    Other approaches to algorithmic fairness include demographic parity and equalized odds, amongst many others \cite{barocas2017fairness}. It is also very well known that the above three crucial measures are at odds with one another \cite{chouldechova2017fair, hardt2016equality, kleinbergFairnessTradeoff, hsu2022pushing}. While there is not yet an industry consensus, our approach to choosing predictive parity is consistent with the general standard for AI fairness outlined in Microsoft’s Responsible AI Standard \cite{MSFT_RAI_Standard}.
\end{remark}

\item When AI does not allocate scarce opportunities, “equal AI treatment” by default means “standard quality of service” and is mathematically defined as a comparison of relevant model performance and product metrics computed across the identified demographic groups to an absolute minimum standard. Product examples include members viewing job recommendations or searching for jobs, members viewing content on their feed, viewing other members to connect to grow their network, etc. See Figure \ref{fig:two-sided} for more details. 
\end{itemize}

%as predictive parity, meaning, the expected outcome of a prediction (e.g., click or no click, invite or not) is independent of the demographic group membership conditional on the score of the prediction (even if this means that different groups are impacted differently in aggregate). 

In adopting this definition, we recognize that “equal AI treatment” does not focus on, or guarantee, equitable outcomes as detailed in Section \ref{sec:intro}. That being said, we are still choosing this as our starting point to be able to clearly disentangle the equal treatment approach from equitable outcomes. We understand that choosing this measure can potentially only maintain the status quo in society, and can even be unintentionally detrimental to underrepresented populations, i.e. members who were not getting equitable outcomes. But that is a \textit{conscious choice}. Maintaining equal AI treatment as defined above ensures that we are not compounding an inherent bias in society with our algorithms. 
This separation is crucial for solving the much larger societal challenge. If our algorithms enforce equal AI treatment, we can bring forth product changes toward achieving equitable outcomes, while ensuring the AI component isn't introducing any unfair bias. We discuss strategies around this in detail in the next section.

We also recognize that for a global service like LinkedIn, there are an unlimited number of demographic groupings (such as race/ethnicity, gender, disability, religion, …) and sub-groupings (e.g. specific racial or ethnic identities) that could be leveraged for delivering “equal AI treatment,” and significant challenges to address, including a lack of accurate data on membership in relevant groupings, regional differences in group definitions or prioritization, evolving norms of group identity, and the difficulty of harmonizing efforts to apply equal treatment across a wide range of interconnected groupings. We don’t view these challenges as a reason not to act based on the data available, but we expect to expand and develop our approach to demographic groupings over time. As a first step, we have launched the Self-ID initiative \cite{selfid1, selfid2} to be able to collect such information (see Figure \ref{fig:selfid} in the appendix) in an ongoing manner in order to make LinkedIn a more equitable platform.  

\begin{remark}
Note that AI can also be used \textit{without} the context of a member for example in generative models or text-based content moderation. In such situations, the mathematical definition of equal AI treatment can differ and be highly context specific. We do not go into such details here due to brevity. Please see \cite{si2022prompting, jalal2021fairness, tan2020improving} and \cite{garg2019counterfactual} for details.
\end{remark}

\subsubsection{Mitigation via a Justifiability Framework:}
For non-demographic groups, if a model is found to be performing poorly across particular groups, the standard approach is to simply add the particular group membership as a feature. %For example, if a model performs poorly for less active members compared to more active members, then a feature for member activity levels could be added. 
However, there are multiple reasons why this approach may not be appropriate or even feasible when considering sensitive demographic data like binary gender, including privacy concerns about data access and usage, and possible unintended consequences of including demographic information in models. Thus even if it is mathematically possible to define mitigation strategies for enforcing predictive parity  \cite{diciccio2022predictive}, we must take a cautious approach since from a legal and policy perspective \cite{xiang2019legal, wachter2021fairness}, there aren’t clear processes to follow that rely on the use of demographic data.  

%We at LinkedIn have and will continue to invest in learning and building out appropriate processes that we can share with others. 
We present a justifiability framework that we propose to follow to determine whether there are means to mitigate algorithmic bias, first, without using demographic information. If not, then we deem that the use of demographic information is justifiable (albeit subject to appropriate guardrails, as a last resort, and only if the benefits outweigh the risks). The framework is not meant to be exhaustive or necessarily include theoretical guarantees; instead, the goal is to create a standard set of questions to ask and investigations to be performed. %Importantly, we designed this process to balance engineering effort and comprehensiveness, as it needs to respect both the availability of the fairness auditors and the engineers of the original product. 
The major steps of the framework are highlighted below:

\begin{enumerate}[noitemsep,topsep=0pt,parsep=0pt,partopsep=0pt]
\item \textbf{Root-Cause Analysis:} Preparing a comprehensive root cause analysis checklist to determine, to the extent feasible, why we observed a prediction parity gap or a difference in the quality of service. We do not want to blindly mitigate the situation without understanding why it happened in the first place. For some initial work in this area please see \cite{ghosh2022faircanary, alikhademi2021can}. 
\item \textbf{Mitigation strategies without demographic information:} We consider alternative equal treatment implementations without using demographic information that results in a smaller predictive parity gap or difference in the quality of service. Examples include collecting more and higher quality training data, for example, to create a more balanced representation of diverse groups \cite{iosifidis2019adafair, han2021balancing}, training larger capacity models, or finding relevant features that are confounding factors of demographic group membership, etc. Please see \cite{hashimoto2018fairness, lahoti2020fairness} and the references therein for more state-of-the-art methods in reducing bias without demographic information. 
\item \textbf{Mitigation using demographic information:} In many instances it may not be possible to close the gap without using demographic information \cite{dwork2012fairness, andrus2021we}. In such cases, we adopt a least granular intervention approach where mitigation requires using demographic data at prediction time (e.g. to calibrate a model \cite{pleiss2017fairness, bella2010calibration, hebert2018multicalibration}). Accordingly, we will generally prefer post-processing \cite{kim2019multiaccuracy, lohia2019bias, petersen2021post} to in-processing \cite{wan2022processing} approaches due to their interpretability and accountability of what exactly is happening under the hood.
\item \textbf{Unintended consequences:} We will assess potential product risks of the proposed AI fairness mitigation (especially when using demographic information), and quantify those risks through experimentation \cite{kohavi2013online, saint2020fairness, friedberg2022representation}. Only when the benefits clearly outweigh the risks will we implement the mitigation for all members of the platform.
\item \textbf{Mitigation Guardrails:} Mitigations to achieve equal AI treatment will not involve affirmative action or interventions seeking equitable outcomes for particular demographic groups. We consider such interventions as better suited to product equity initiatives for several reasons.
\begin{enumerate}[label=(\alph*)]
\item Pursuing equitable outcomes requires qualitative user research, a deep understanding of product features and how they affect different demographic groups, assessment of benefits and risks, and close governance and oversight.
\item AI models can change with time and can suffer from a lack of monitoring or maintenance, leading to potential unintended consequences of affirmative action interventions within a model.
\item The legal and social context in which a product exists–as well as the equity challenges and priorities of a product–can evolve over time, demanding that any affirmative actions at an AI level be kept fully synchronized with those changes, which represents a significant operational challenge.
\end{enumerate}
\end{enumerate}
We highlight each of these mitigation steps in detail in Section \ref{sec:casestudy} where we deep-dive into a particular case study.

\subsubsection{Privacy and Security:}
\begin{figure*}[!h]
    \centering
    \includegraphics[width=0.8\textwidth]{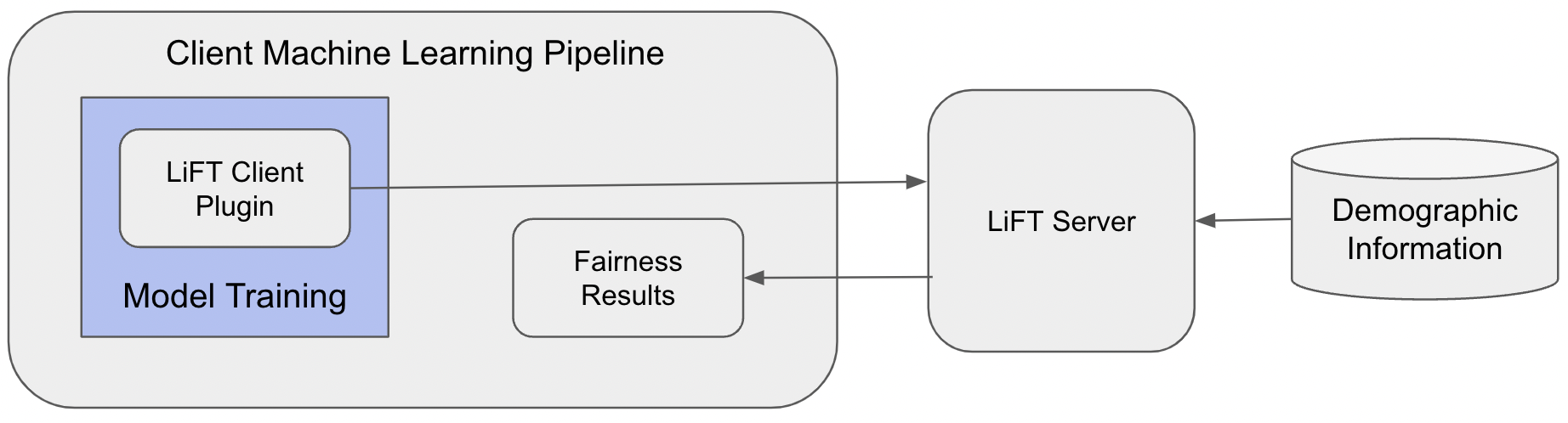}
    \vspace{-0.3cm}
    \caption{Client-Server architecture which allows a model owner to evaluate the fairness metrics for their model without needing access to demographic information data. LiFT here refers to the LinkedIn Fairness Toolkit \cite{vasudevan2020lift} which houses the measurement algorithms.}
    \label{fig:client-server}
\end{figure*}
One of the major steps in being able to accurately measure and potentially mitigate algorithmic bias relies on access to accurate group information. Although we at LinkedIn began asking for gender and disability demographic data globally many years ago, we realized that it did not encompass the broad spectrum of identities of our members. In 2021, we introduced the “Self-ID” initiative to expand on those efforts in the U.S. \cite{selfid1, selfid2}. 

When members choose to provide this data, we handle this data under strict privacy and security measures. By default and unless they have a legitimate business need, LinkedIn engineers do not have access to this data in an unencrypted format. However, the same protections used to safeguard this sensitive data from unauthorized use or disclosure, also create obstacles in using it for measurement of algorithmic bias across our relevant products. To address this challenge, new and innovative work is underway both from an engineering and an algorithmic aspect to be able to leverage this data without undue risk. Initiatives such as privacy-preserving machine learning \cite{mohassel2017secureml, al2019privacy, xu2015privacy}, homomorphic encryption \cite{fontaine2007survey, yi2014homomorphic, sun2018private}, A/B testing under differential privacy \cite{friedberg2022privacy}, etc., are all being studied across various teams. 

Given one of our focus areas is group-based measurement, it may be possible to use differential privacy \cite{fioretto2022differential, dwork2014algorithmic, dwork2008differential} to obfuscate the true group membership information, but still be able to estimate predictive parity differences with some degree of accuracy. We can mathematically derive an unbiased estimate of the fairness metric by knowing the noise that is being added to the group information data \cite{personalizedads2023, fioretto2022differential}. Moreover, for an extra layer of protection, none of the product-focused AI teams would need access to this data for evaluating their models. We have developed the pipelines through a server-client architecture (see Figure \ref{fig:client-server}) such that each model owner would be able to evaluate their model for fairness without ever needing access to this information \cite{fairnessblog1}.

Although it might be possible to use demographic data for measurement purposes, the problem is much more complex and nuanced for mitigation. The safest mitigation technique is when we are able to mitigate a predictive parity gap without needing access to demographic attributes. However, in instances where that is not possible, we do plan to use demographic information in order to mitigate our models, albeit under limited use cases and with protections designed to ensure limited access for this use case. 

An obvious question arises, why not use obfuscated information (similar to the differentially private approach for measurement) in doing mitigation? The answer to that lies in being very conscious about our choice. Our post-processing mitigation solution \cite{diciccio2022predictive} for enforcing predictive parity changes the score based on group membership. In the case where we are using noisy group information data, we would know for certain that for those fraction of noisy cases, we have changed their score in a wrongful manner. At the end of the day, these score changes happen at an individual member level and we strongly believe in the fact that \textbf{we should not knowingly give a wrong score leading to undeserving opportunities}.

Overall, operationalizing this framework requires various engineering aspects to ensure that we adhere to the relevant laws and our privacy commitments, including but not limited to internal controls regarding employees who have access, reducing access to only when necessary, safeguards to prevent data misuse and system infrastructure such that equity can only be achieved without giving internal stakeholders direct access to such private information.

\begin{remark}
Data availability: We may not have the demographic data for many of our members, which makes it difficult to apply the post-processing techniques. We are currently researching new methods that may work without the strong dependency on demographic data. For details please see \cite{brian2023survey}.
\end{remark}

\subsection{Principle 2: We will not consider equal AI treatment the end of our work, but will treat it as the foundation of a broader fairness and equity strategy.}

We consider equal AI treatment, as a necessary foundation, but not itself sufficient for achieving equitable product experiences or outcomes. For example, equal AI treatment would not address a group’s under-representation in recruiter search results or connection recommendation if that under-representation stems from a real-world under-representation in a given sector. Achieving the broader equity goals requires that equal AI treatment be complemented with a product equity strategy and in partnership with customers who choose to pursue equity goals. Other measures – such as product features and design changes – are outside the scope of AI initiatives and require different inputs and assessments.% (e.g., qualitative user research, social science learnings, customer requirements, and effects on different demographic groups).

\begin{figure*}[!ht]
    \centering
    \includegraphics[width=\textwidth]{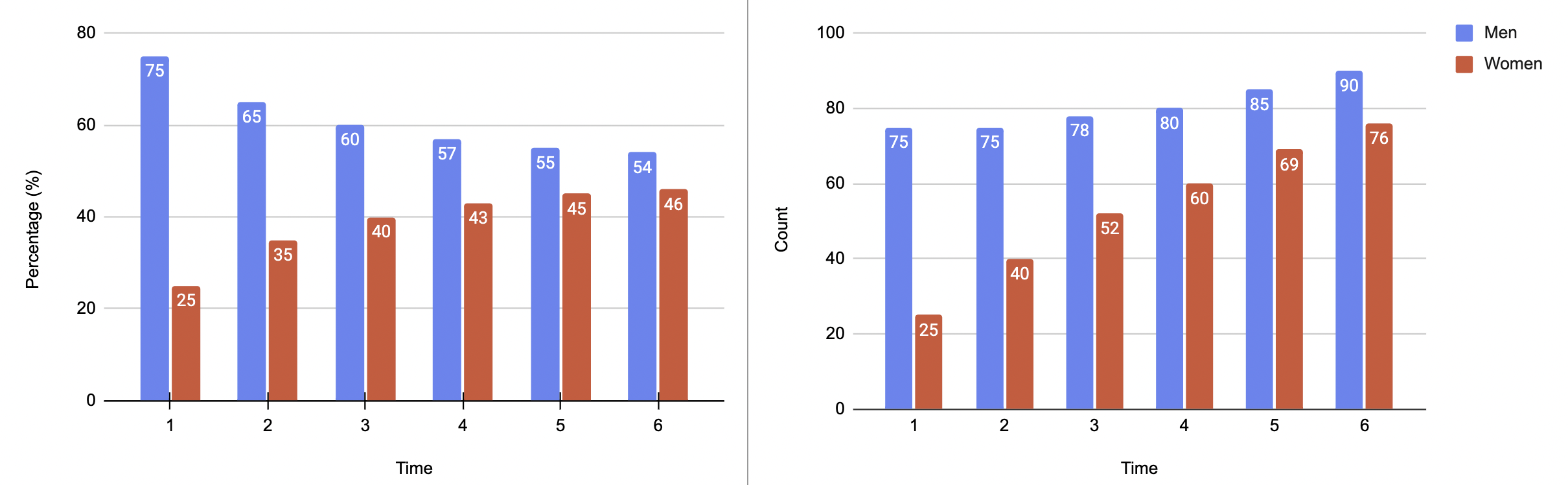}
    \caption{A hypothetical situation over time where we start with an initial distribution of 75\% and 25\% and through exploration ultimately converge to 54\% and 46\% (Figure on left). Note that in many situations it may not be possible to reach an absolute equal distribution. At the same time, we are never sacrificing on the total number. The figure on the right shows how their outcomes are growing over time thereby bringing us to a more equitable future. Moreover, throughout this time frame, we are ensuring that equal AI treatment holds for the members in situations where we do not have active exploration.}
    \label{fig:ee}
\end{figure*}

We have already started our journey toward achieving equitable product experiences. The first step of this journey undoubtedly starts with understanding how our members experience our platform, especially those from historically and systemically marginalized communities. Through the Self-ID initiative, more than 10 million members \cite{selfid3} have shared some aspect of their identity on LinkedIn. As more members join us on this equity journey and Self-ID, we’ll be able to launch new products and experiences to help drive more equitable outcomes for those members facing barriers. One such product feature that we have already shipped is ``diversity nudges'' in LinkedIn Recruiter and LinkedIn Learning \cite{diversityNudges}. These features allow recruiters and hiring managers to increase gender representation in their qualified candidate pools. We also have a feature to hide names and photos of candidates to reduce unconscious bias \cite{hideNames}.

Beyond having a product equity strategy, it is also crucial to understand the potential unintended consequences that interventions aimed at ensuring equal AI treatment can have. Given how tightly interconnected AI and product outcomes are, we take an end-to-end view. Any AI change, although only targeted at ensuring equal AI treatment, can have a strong impact on product metrics. As a result, it is necessary for product teams to commit to supporting equal AI treatment mitigation efforts and monitoring outcomes against product equity strategy and goals over time. For example, addressing a prediction gap may result in increased visibility or participation for a given group (e.g. female profiles) with attendant risks of harassment which must be assessed.

\subsubsection{Towards Equitable Outcomes via AI} Note that although we stated that AI systems are expected by default to meet equal AI treatment, it is also possible to devise AI changes with the explicit aim of achieving equitable outcomes. Such interventions are thought of as product design changes that leverage AI rather than direct AI interventions aimed at equitable outcomes. An example would be a product that reserves a portion of the screen real estate to show underrepresented people or content, using AI to retrieve the highest ranked members or items of that group. This is AI serving a product design decision driven by an equity strategy, rather than a modification of the AI algorithm to satisfy an equity strategy. This conceptual separation is crucial to be able to structure and adhere to the principles. 

As we mentioned earlier, the notion of equal AI treatment by enforcing predictive parity can reflect and doesn’t disrupt the status quo in society. It is a valid criticism that in a world where predictive parity holds, we might not be giving the deserving opportunity to some members only because their group in the past, did not engage with the platform. More formally, if there are groups of individuals who rarely visited or engaged with the platform (thereby mostly having $Y = NA$), enforcing predictive parity would give them a very low score ($\hat{Y} \approx 0$) and hence they would either be very low in the ranked list or not shown at all. As a result, if there is a member who is highly qualified (with a potential of getting $Y = 1$)\footnote{Unless specifically mentioned we are usually working in the regime of binary classification.}, but belonged to this group, enforcing predictive parity will not give the deserving opportunity to this member. 

We believe such situations should be solved through an explore/exploit strategy \cite{agarwal2009explore, liu2018explore} as part of the product equity strategy and goes beyond the equal AI treatment notion. Consider a situation in time, where our models have enforced predictive parity. We now do not know if there are members who might be qualified but are not being shown due to our enforcement of predictive parity. At this stage, we consider a budget for exploration. Let’s assume that now for 95\% of all member sessions, we are maintaining predictive parity, while for 5\% of sessions, we are actively trying to explore members who we think might be qualified, but we lack the data to accurately predict it. There are several ways of doing this, for details please see \cite{chouldechova2020snapshot, joseph2016fairness, jabbari2017fairness}. Based on this exploration data, if we can truly find some of the missing qualified members, then their data is automatically passed into the training model and over time, the model understands that these are qualified members and automatically shows them without the need for artificial exploration. %Similarly, we understand unqualified members and the model learns to rank them lower. 
Overall through these iterations, the main goal is to maintain equal AI treatment for the majority of the sessions (say 95\%) and slowly improve the actual metrics toward achieving equitable outcomes. 

As a concrete example, consider a case where an AI researcher sees 75\% men and 25\% women in their connection recommendations and the ranking algorithm satisfies predictive parity. Through the explore/exploit paradigm, our aim would be to improve this and potentially make it  65\% men and 35\% female or even better in the long run, while maintaining predictive parity. 
Figure \ref{fig:ee} (on the left) shows the cycle of improvement over time. It is important to note that in the above example, recommended members being viewed by this AI researcher meet the feature criteria to be recommended. Thus, we are never replacing a qualified member with someone who is unqualified. Moreover, we are only working with percentages above and not raw counts. In an ideal situation, we would expect the total count to grow over time. For example, if we initially started with 75 men and 25 women, we can call our strategy a success if we now have 75 men and 40 women, raising the total count but reducing the percentage gap. See Figure \ref{fig:ee} (right).

We believe that tracking these proportional and count statistics can give insight into the long-term effects of the explore/exploit strategy and indicate if we are reinforcing the status quo or moving towards a more equitable environment. Still, there are potential pitfalls in this strategy that our random sampling-based exploration does not explicitly consider. For example, it is possible that the underrepresented group is simply harder to learn or lacks samples (even with randomization). Hence, if we reach a male-to-female 70/30 steady-state proportion, it is not clear if this is the true long-term behavior or if it is because a model lacks the capacity to adapt to the newly gathered samples. One method for addressing this could involve ranking based on confidence intervals of the expected outcomes $E[Y|S]$ rather than the point estimates as suggested by \citet{salem2019secretary}, which could better account for model uncertainty and the potential for a candidate to rank higher, rather than the point estimate. In a similar vein, bandit-based approaches that utilize the confidence intervals can pose a more targeted strategy than random sampling. Lastly, if there are a-priori contextual reasons that a group should have higher outcomes than they currently do, it is also possible to empirically test this hypothesis by incrementally boosting the groups’ scores and running a post-hoc analysis to determine if they have been under-scored. Learning under-selection bias remains an open research problem.

Overall, such AI strategies, along with specific product launches are also part of our overall equity strategy toward achieving equitable outcomes. The prior work on generating a representative list in talent search \cite{geyik2018talent} also falls into this framework as an example of an AI intervention targeted towards equitable outcomes, where the authors tried to match the underlying gender distribution of a search query to the ranked results.

\subsection{Principle 3: We will validate our approach externally and lead with transparency.}

AI fairness is a developing field where LinkedIn is well-positioned to make significant contributions. We aspire to help and inspire other organizations as they seek to deliver equal AI treatment and explore broader equity strategies. We will share our learnings through case studies, articles, and other presentations, and we will leverage input and collaboration from our members, our customers, academia, industry partners, and advocacy groups, as well as Microsoft’s RAI and social science resources. We will explore approaches to algorithmic auditing by vetted external expert groups with the goal of delivering an external quantitative assessment of our measurement and mitigation approach.

\section{Case-Study}
\label{sec:casestudy}

This section illustrates how the methods described in Section \ref{sec:practice} are applied in a real-world example. We focus on a people recommendation product at LinkedIn called “People Also Viewed” (PAV, see figure \ref{fig:browsemaps} and the appendix). This product recommends additional profiles you might be interested to connect with or learn from based on the current profile you are viewing and is an example of the scarce resource allocation paradigm (limited number of recommended profile slots). The product is the second largest source of traffic for profile views, therefore any negative bias (failing to surface qualified recommendations) could adversely affect members’ abilities to grow their network. On the other hand, ‘over-recommending’ members can lead to harm such as unwanted connection requests and spam. 
In this case study we focus on measuring and mitigating bias for binary gender. Equal AI treatment in this case means that the model that recommends related profiles for PAV produces ranking scores that satisfy predictive parity across binary gender.

\vspace{-0.1cm}
\subsection{Measurement}
\begin{figure}[!ht]
    \centering
     \begin{subfigure}[b]{0.5\textwidth}
         \centering
         \includegraphics[width=0.5\linewidth]{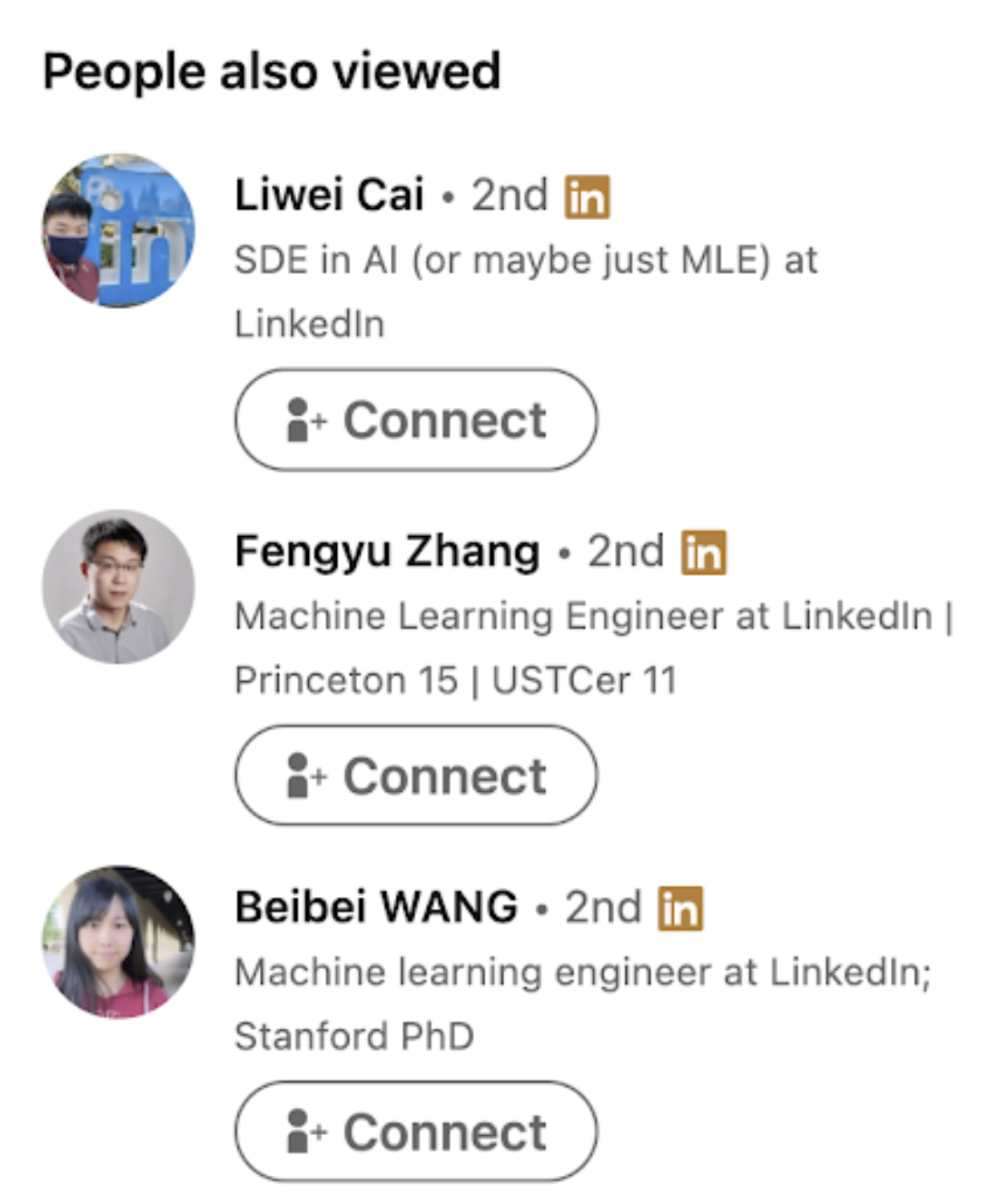}
         \newsubcap{The People Also Viewed product that shows a ranked list of recommended members.}
        \label{fig:browsemaps}
     \end{subfigure}
     \hspace{1cm}
     \begin{subfigure}[b]{0.5\textwidth}
         \centering
        \includegraphics[width=0.9\linewidth]{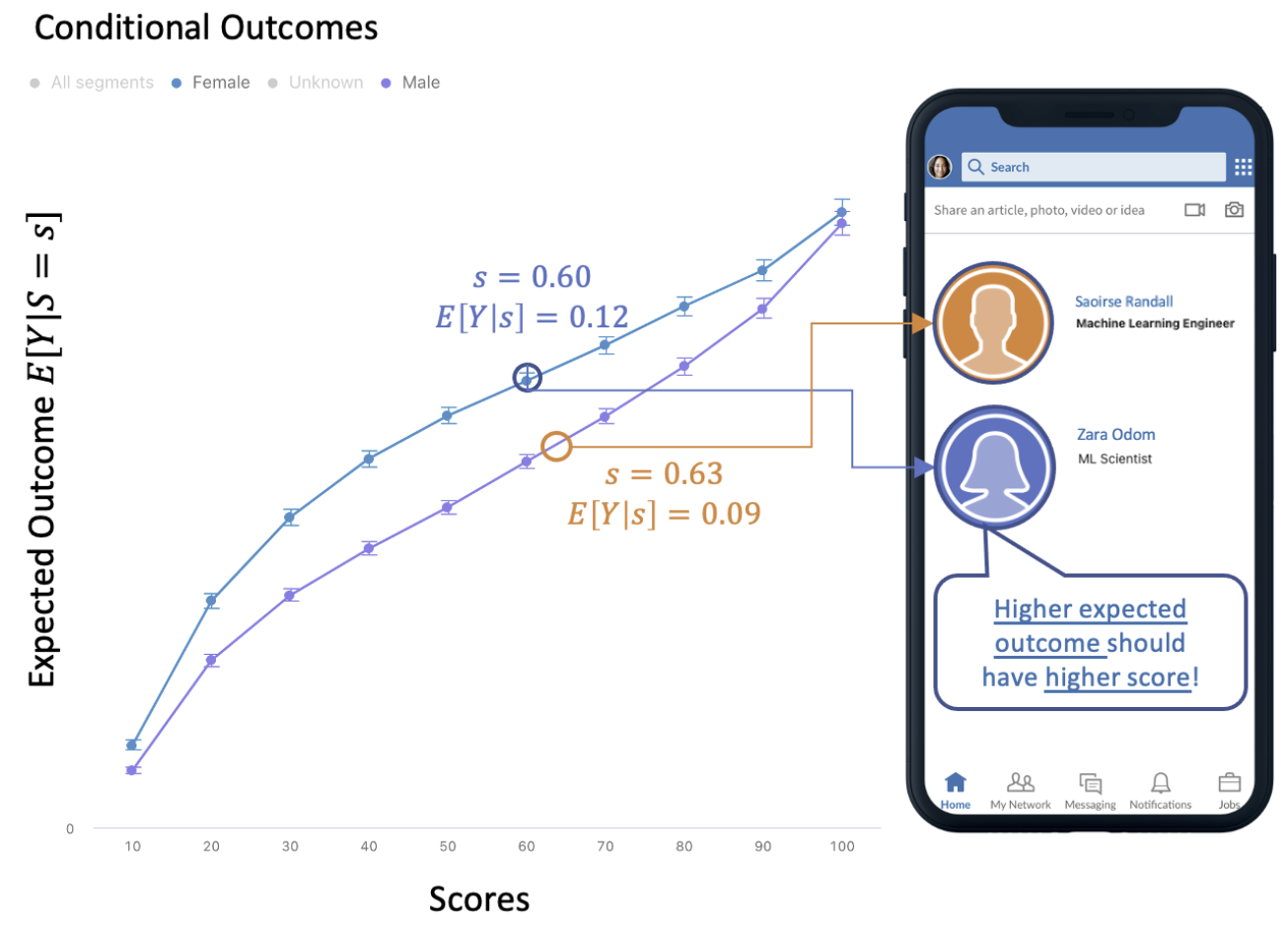}
        \newsubcap{How predictive parity gap changes the ranking order in products such as People Also Viewed.}
        \label{fig:pavPP}
     \end{subfigure}
\end{figure}

The motivation for this use case is an observed gap in predictive parity between male and female members\footnote{We did not observe a viewer-side gap in this product since the model does not use viewer-side features.}. Figure \ref{fig:pavPP} illustrates the real-world impact of lack of predictive parity. Consider a group of female and male members that share a similar AI relevance score: the females in that group will see higher real-world outcomes of actual network formation. This means that the score should have been higher for the females in that group. The model is therefore under-predicting for females. Meeting equal AI treatment would require the two curves in \ref{fig:pavPP} to overlap.

%To illustrate why gaps in these conditional expected outcomes ($E[Y|\hat{Y}=s, A=a]$) lead to fairness concerns in PAV, consider the scenario illustrated in Figure \ref{fig:pavPP}. Differences in expected conditional outcome across groups can cause one group to have an unfair advantage, where they can be ranked above a member of the other group despite having lower “qualification” (expected outcome). 
%Models can vary in the predictive parity gaps as they are trained and evaluated across different data. %We have developed tools within our scalable machine learning framework that enable developers to assess the fairness properties of their models during training similarly to performance metrics \cite{fairnessblog1}. 

The remainder of this Section details our approach to mitigating this predictive parity gap. Although we focus on binary gender, the framework can be extended to other demographic groups, assuming there is sufficient data.

%\vspace{-0.2cm}
\subsection{Mitigation via Justifiability Framework}
We follow the steps in Section \ref{sec:practice} to mitigate the observed gap in predictive parity.

\subsubsection{Root-Cause Analysis}
%A common question whenever bias is identified is, why? 
Our framework provides a standard set of investigations that can provide insights into the mechanisms leading to a (gender) gap \footnote{Note that these methodologies can be applied to general model improvement, and are not limited to just closing a fairness gap.}. %Some steps are meant to identify (or rule out) obvious red flags, while others provide more granular analysis. 
 %This is a benefit of our choice to measure fairness using predictive parity, as we assume that improving demographic attribute-specific calibration will directly improve model performance \cite{durfee2022heterogeneous}. In other words, training a fair model does not need to come at the expense of actual model performance \cite{shah2022selective}.

\begin{enumerate}%[noitemsep,topsep=0pt,parsep=0pt,partopsep=0pt,wide, labelwidth=!, labelindent=0pt]
    
\item \textbf{Distribution of binary gender:} The obvious first step is breaking down the data to see the ratio of binary gender; in the most trivial case, if one gender is missing from the data, it would explain why the model is miscalibrated for this group. In the less extreme case, binary gender skew could still contribute to bias if the relevant signal for one gender is not large enough for the model to learn predictions for each gender. In the case of PAV, we did not see a large gender skew for the source or destination members (see Table \ref{tab:genderdist}).\footnote{There was a 29pp difference between men and women for the viewer gender, but given the model did not use viewer features at all, we did not see mitigate on that.}.

%binary gender skew was greatest for viewing members (29pp difference between men and women), which are the members who determine the relevance of the model’s recommendations . However, the model only uses features related to the current profile member and the recommended member. The lack of features for such a skewed group, particularly for the group actually generating labels for the model, could lead to suboptimal recommendations for the underrepresented gender.
\begin{table}[!htp]\centering
\begin{tabular}{lrrr}\toprule
&Men &Women \\\midrule
%Viewer &67.90\% &32.10\% \\
Source &54.50\% &45.50\% \\
Destination &54.70\% &45.30\% \\
\bottomrule
\end{tabular}
\caption{The distribution of gender across different member groups. The results are normalized to binary gender. Unknown, non-disclosure and non-binary gender are not shown.}\label{tab:genderdist}
%\vspace{-0.3cm}
\end{table}

\item \textbf{Feature justifiability:} Building AI models responsibly requires that we only include features if they are justifiable. 
Although we do not give a formal definition, the use of a feature is justified, for example, if the feature is relevant to the modeling task. Domain knowledge and common sense are required to judge that relevance in context.

The justifiability bar is highest for demographic information or other highly confidential data. In addition to privacy and security concerns, including these data in a model could exacerbate biases or lead to other unintended consequences. This is why we ask that the unjustifiable features be removed even if model performance drops. Removing a feature that isn't justified doesn't mean all other features correlated with it also need to be removed. On the contrary, as we discuss in Section \ref{sec:mitigationWOPA}, one way to close the predictive parity gap is to add relevant correlated features on a causal chain. An audit of the PAV model features did not reveal any unjustifiable features.

%Note that justifiable features (e.g. click behavior patterns) can sometimes be correlated with sensitive demographic feature. Let's take gender as an example. We do not ask for additional justification for correlated features because a) it is unreasonable to expect that all features are uncorrelated with gender, and b) adding justifiable features correlated with gender can actually close the predictive parity gap. See Section \ref{sec:mitigationWOPA} for more details. An audit of the PAV model features did not reveal any unjustifiable features.

\item \textbf{Cohort-level Error Analysis:} We have observed that PAV is miscalibrated when segmenting on binary gender. However, the metric is a population-level average, meaning we do not have insight into the level of miscalibration for different subgroups. To get a more granular understanding of what is causing this binary gender-based gap in predictive performance, we perform a cohort-level analysis. 

We build a tree model to split the data using a user-defined error metric. The splits are determined such that the partitions of the data maximize the differences in the error metric. The resulting cohorts and the features used in the splits can give more granular insight into model performance compared to standard feature importance methods \cite{hooker2018evaluating}, as the data is automatically segmented into high and low-error cohorts. Furthermore, we can include non-model features as candidates in the cohort splits, and we can also use a non-model error metric as the splitting criterion \cite{xu2019explainable}.

For the fairness use case, we set the metric as the residual between the label and the predicted score. This is a proxy for miscalibration error, which is essentially what predictive parity tries to measure. As the goal is human-interpretable cohorts, we limit the depth of the tree to be three, so that only eight cohorts are generated. For candidate features, we include the gender of any member in the model (for PAV, these would be the viewer, current profile member, and recommended member) in addition to the actual model features. Figure \ref{fig:errorCohortBrowsemaps} (see appendix) shows the result of running the error cohort analyzer on PAV data. The error cohort model used the binary gender of both the recommended member and the current profile member as top feature splits, with MALE cohorts generally having lower residuals compared to FEMALE cohorts. This suggests binary gender could be a driving factor for the measured bias, increasing the likelihood that gender will need to be directly used for mitigation.
\end{enumerate}

\subsubsection{Mitigation experiments without demographic information}
\label{sec:mitigationWOPA}

Mitigation of fairness violations is typically studied in contexts that allow access to demographic information during model training and inference. However, a key component of our justifiability framework is our proposition that demographic data should only be used when other methods are demonstrably inadequate, and when we were unable to find negative unintended consequences from mitigating using demographic data. We therefore evaluated a wide range of methods for mitigation without demographic data at inference time, from feature-selection methods to fairness-constrained in-processing methods. This section provides an overview of these techniques, and presents our rationale for ultimately choosing to mitigate with demographic data. 

The problem of closing the gender calibration gap without using gender as a feature can be cast as one of identifying missing features that are relevant and correlate with gender. We motivate this approach by visualizing gender and model features in a causal graph \cite{pearl2009causality}, where the terminal node is model prediction. Our goal is then to determine if there are additional, non-gender features that could ‘block’ the effects of gender on the gap such that the gap is reduced.

\begin{figure}[!ht]
     \centering
     \begin{subfigure}[b]{0.4\textwidth}
         \centering
         \includegraphics[width=0.8\textwidth]{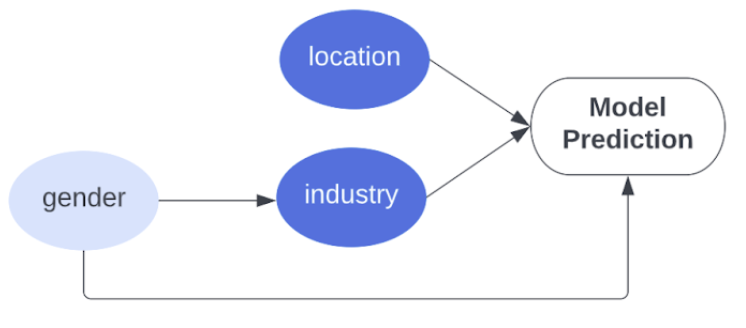}
         \newsubcap{Original model: Gender affects model prediction both through a direct path as well as an indirect path through industry}
         \label{fig:dag_left}
     \end{subfigure}
     \hfill
     \begin{subfigure}[b]{0.48\textwidth}
         \centering
         \includegraphics[width=0.8\textwidth]{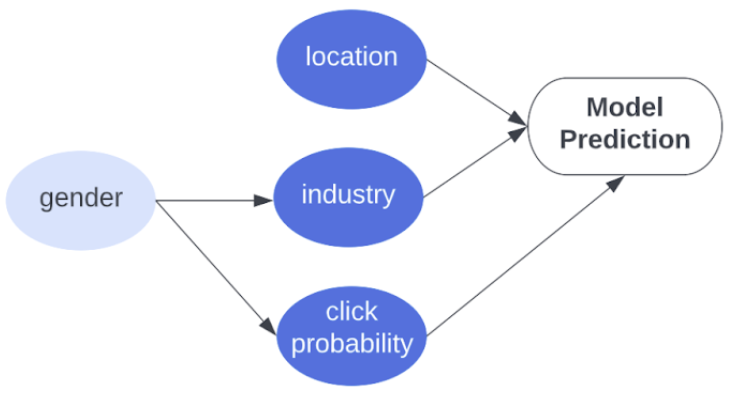}
         \newsubcap{New model: Click probability identified as a new feature and Gender affects model prediction only through indirect paths.}
         \label{fig:dag_right}
     \end{subfigure}
\end{figure}

Let’s consider a toy model with only two features; we show the causal relationships between the features, model predictions, and binary gender (which is not included as a feature) in a directed acyclic graph (DAG) (see Figure \ref{fig:dag_left}). The original model is biased because binary gender is causally related to the predictions through a direct path. Imagine we identify a new feature, click probability, that is a descendant of binary gender and fully captures its effect. When click probability is included in an updated model version, binary gender no longer has a direct effect on the model predictions (see Figure \ref{fig:dag_right}). We can therefore theoretically close any gender-based bias by including missing features in the direct causal path of binary gender.
A caveat of this approach is that our toy example only shows the ideal case where the new feature closes the gap and does not have additional interaction effects. In reality, any new node in a large graph can drastically change the DAG by adding new edges and therefore changing causal relationships \cite{guo2020survey}. New features in the direct path of binary gender may even widen the gap, therefore we cannot indiscriminately add features we think are correlated with binary gender and outcome.

One approach for a more intentional mitigation strategy is to leverage domain expertise to identify relevant missing features; however, it is difficult to scale these ad-hoc solutions. We therefore first created a superset of common LinkedIn member features and tested automated feature selection algorithms to pick the subset that could close the gap. The two main strategies we tested were Quantile Prediction Drift (QPD) and a method based on causal effect decomposition \cite{chakrabortty2018inference}. We also tested more ‘standard’ techniques such as brute force search of the feature space and imputation of missing feature values (identified as another possible issue from root cause analysis). Finally, we ran experiments including binary gender in the model to get a direct comparison between gender-aware and gender-blind techniques. The results are outlined in Table \ref{tab:mitigation_results} and we observe that adding generic LinkedIn features showed no reduction in the fairness gap, regardless of the feature selection strategy. Gender-aware approaches consistently outperformed gender-blind options to close the gap.

\begin{table*}[!htp]\centering
\begin{tabular}{lccc}\toprule
\textbf{Mitigation strategy} &\textbf{AUROC (\%)} &\textbf{Predictive Parity} &\textbf{Reduction from} \\
\textbf{} &\textbf{} &\textbf{Gap (\%)} &\textbf{baseline (pp)} \\\midrule
Post-Processing gender-based calibration (offline) \cite{diciccio2022predictive} &\textbf{66.3} &\textbf{0} &\textbf{13.3} \\
Add gender* &65 &2.8 &10.5 \\
New feature superset + gender* &66.8 &4.6 &8.7 \\
Imputation with gender-specific means* &65 &8.4 &4.8 \\
QPD selection** &65.5 &10.7 &2.6 \\
Causal effect decomposition** &65.2 &11.8 &1.5 \\
New feature superset &66.8 &13.1 &0.18 \\
Imputation with population means &64.8 &13.2 &0.05 \\
Baseline &65.1 &13.3 &NA \\
\bottomrule
\end{tabular}
\caption{A condensed summary of mitigation experiments. *Not a justifiable mitigation strategy, but useful for offline comparison.
**Best combination of features
}\label{tab:mitigation_results}
\end{table*}

While we have focused on data-based mitigation methods, there are also other avenues of mitigation. Some examples include tuning model parameters to optimize for calibration, training with calibration-motivated loss functions, or post-processing with group-agnostic calibrators. As part of our efforts to identify effective demographic data-restricted methods for mitigation, we conducted a wide survey of methods that varied in both the point of intervention (pre-/in-/post-processing) as well as the volume of demographic data required (e.g., only in training vs. in training and inference). We present this survey in a separate paper, see \citet{brian2023survey}, but note here that the overwhelming conclusion is that using gender in post-processes was by far the most effective strategy for achieving predictive parity fairness.

Given that our thorough experimentation with gender-blind and gender-limited techniques did not lead to successful mitigation, we felt \textbf{justified in using the post-processing gender-based calibration} to mitigate the algorithmic bias in PAV.

\subsubsection{Mitigation experiments with Demographic Information}
We use the post-processing technique detailed in \citet{diciccio2022predictive} as our bias mitigation training (BMT) algorithm which requires the use of  demographic information. Intuitively, BMT fulfills predictive parity by setting $E[Y|\hat{Y}=s, A=a]=s$ for all groups $A$ (here $s$ is the predicted score from the model). As shown in Table \ref{tab:mitigation_results}, BMT (row 1) successfully reduced the gap in offline experiments.

%BMT is a per-group calibrator, that takes in the predicted score $s$ for an individual from group $g$ and assigns the expected value $E[Y|S=s, G=g]$ with piecewise linear functions \cite{diciccio2022predictive}. In short, BMT fulfills predictive parity by setting $E[Y|S=s, G=g]=s$ for all groups. 

\begin{figure}[!ht]
     \centering
     \begin{subfigure}[b]{0.45\linewidth}
         \centering
         \includegraphics[width=0.9\linewidth]{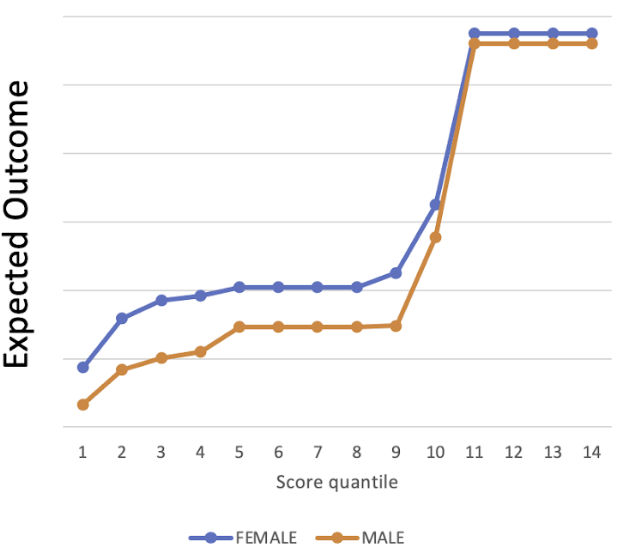}
         \newsubcap{Online results of base model}
         \label{fig:bmt_orig}
     \end{subfigure}
     \hspace{0.1cm}
     \begin{subfigure}[b]{0.45\linewidth}
         \centering
         \includegraphics[width=0.9\linewidth]{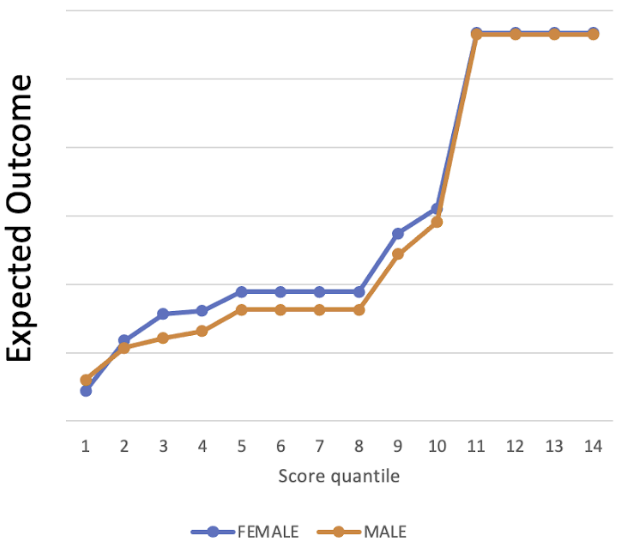}
         \newsubcap{Online results of Bias Mitigated Model.}
         \label{fig:bmt_post}
     \end{subfigure}
\end{figure}

Based on our root cause analysis and offline mitigation experiments, we concluded that BMT was effective (offline) and should be experimented online. We launched an online A/B test comparing the baseline PAV model with a BMT-mitigated version (Figure \ref{fig:bmt_orig} and \ref{fig:bmt_post}) that also showed a reduction in the predictive parity gap. Furthermore, the BMT model showed significant 2\% lift in total profile actions (e.g., clicking the recommended profile), which is the main product metric, confirming our hypothesis that equal AI treatment need not sacrifice general model performance.

%\vspace{-0.3cm}
\subsubsection{Measuring unintended consequences} Ensuring equal AI treatment in our case study requires that the gap between AI scores attributed to males and females be closed. Relative to male scores, female scores will now be higher. The product consequence of this is that a higher proportion of female profiles will be shown on the PAV product. Before we can ship this equal AI treatment intervention, aimed at eliminating predictive bias, we need to make sure there are no negative real-world consequences. One potential risk is that females might now receive more unwanted messages, invites, or even harassment. To quantify this risk, we built edge-level (member $\rightarrow$ female) metrics that measure the rate at which females accept connection request, respond to messages received and report unwanted connection requests. We compared the equal AI treatment mitigation described above ("treatment") to the model before mitigation ("control").\footnote{Technical note: Our mitigation is applied on the viewer side and thus it is possible for a recommended member to have their score treated by BMT in some PAV tabs but not others. This can lead to two-way interference effects in the measurement of edge-level metrics such as response rate. Although methods exist for AB testing in two-sided marketplaces such as \cite{nandy2020b, nandy2021b}, they are an open area of research and require highly customized online experimentation implementations. Hence, while the edge-level analysis we proposed and conducted does not rigorously account for two-way interactions, we feel that it provides a reasonable heuristic for gauging these treatment effects.}

%We perform an edge-level analysis (detailed in the appendix) to compare the effects of the treatment member $\rightarrow$ female member and control member $\rightarrow$ female member to determine if there are substantial differences. We focus on females because they will be surfaced more than before in the mitigated model. By measuring edge-level metrics, we can assess whether members recommended by the mitigated model may be experiencing negative effects compared to control \footnote{Technical note: Our mitigation is applied on the viewer side and thus it is possible for a recommended member to have their score treated by BMT in some PAV tabs but not others. This can lead to two-way interference effects in the measurement of edge-level metrics such as response rate. Although methods exist for AB testing in two-sided marketplaces such as \cite{nandy2020b, nandy2021b}, they are an open area of research and require highly customized online experimentation implementations. Hence, while the edge-level analysis we proposed and conducted does not rigorously account for two-way interactions, we feel that it provides a reasonable heuristic for gauging these treatment effects.}. Note that in addition to these new metrics, we also require that the original PAV metrics do not show performance drops, as we do not want to launch a fair but poorly performing model. The actual results are highlighted in the Table \ref{tab:edge_small}.

\begin{table}[!ht]\centering
\begin{tabular}{lrrrrr}\toprule
\textbf{Metric} &\textbf{ControlRate} &\textbf{TreatmentRate} &\textbf{RelativeDiff} \\\midrule
accept &0.20275 &0.234024 &0.154251 \\
reply &0.312765 &0.306352 &-0.020507 \\
report &4.70e-5 &4.60e-5 &-0.017529 \\
\bottomrule
\end{tabular}
\caption{Edge-level response rates across metrics comparing the base model with the BMT model.}\label{tab:edge_small}
%\vspace{-0.5cm}
\end{table}
 Table \ref{tab:edge_small} summarizes the experimental results. The treatment group accepts connection invitations at a higher rate and reports problematic invitations at a lower rate. This means higher network formation and fewer unwanted connection requests. We observe a slight a decrease in messaging response rates, tolerable given the other two positive outcomes. We therefore \textbf{did not find unintended consequences that arise from ensuring equal AI treatment} for PAV.

%\vspace{-0.2cm}
\subsection{Key takeaways from the Case Study}
We were able to measure and mitigate the predictive parity gap that was observed in our PAV model. Through initial root-cause analysis, we were confident that binary gender was a driving factor in the parity gap that was observed. We tried extensive experimentation to try to close that gap without using gender information, however, none of the gender-blind techniques were successful. This led us to justifiably use gender to close the gap as a last resort through post-processing. Before shipping the mitigation, we measured to ensure the mitigation resulted in the product working better and without unintended consequences for the under-served female group. Fairness is a process, and mitigations need to be reevaluated and revisited periodically going through the steps of the justifiability framework.

%\vspace{-0.2cm}
\section{Discussion}
\label{sec:discussion}

We have presented LinkedIn's framework and principles for equal AI treatment. The framework disentangles the definition of AI fairness and brings operational clarity on what AI practitioners are accountable for. Recognizing that AI fairness isn't only an AI problem, a robust product equity strategy needs to complement equal AI treatment. 

Mitigating AI fairness deficiencies suffers from an additional level of complexity: whether or not it is justifiable to use demographic information to close AI bias gaps, and how precisely to use this data. We have also presented a ``justifiability framework'' that addresses this question. The framework requires a thorough root-cause analysis to understand predictive parity gaps and demands exploring alternative mitigations that don't require using demographic data. If the only way to close the predictive parity gap is to directly use demographic data, the framework asks for the minimal possible intervention and for experimentation to detect possible unintended consequences and to confirm the resulting product outcomes are beneficial, and that these benefits outweigh potential risks.

An apparent paradox emerges: using demographic information for mitigation may seem like no longer treating every group the same. The opposite is true: by using demographic information we ensure the scores an AI gives to members of different groups map equally well to real-world outcomes. Equal AI treatment means equally good scores for every group independent of group membership. 

Finally, fairness is never ``done''. Alternative mitigations that don't require using demographic information should continue to be explored and should be preferred. However, unaddressed bias causes real-world harm, and when considering how to mitigate, we need to take into account the cost of inaction. 

We hope this framework can be leveraged broadly and we encourage other organizations to share and collaborate.

%\begin{acks}
%\end{acks}

\pagebreak
\bibliographystyle{ACM-Reference-Format}
\bibliography{biblio}

\pagebreak
\appendix
\label{sec:appendix}

\section{Error Cohort Analysis}
\begin{figure*}[!hb]
    \centering
    \includegraphics[width=\textwidth]{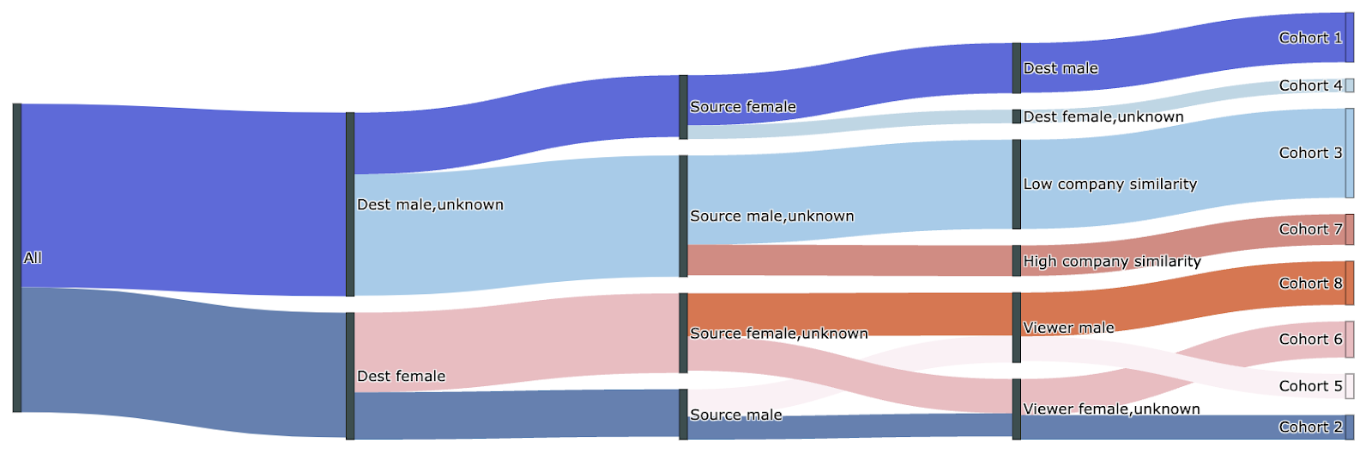}
    \caption{The Sankey diagram for People Also Viewed}
    \label{fig:errorCohortBrowsemaps}
\end{figure*}

The Sankey diagram (Figure \ref{fig:errorCohortBrowsemaps}) shows the flow of members from the entire data (left) to individual cohorts (right). Cohorts are numbered in ascending order based on residual error. For example, cohort 1 has the most negative error (corresponding to more overpredicted samples) and is the brightest blue. On the other hand, cohort 8 has the most positive error (corresponding to more ‘underpredicted’ samples) and is the brightest red. Text describes the features used in the splits.

\section{Edge-Level Analysis for Measuring Unintended Consequences}

\begin{figure*}[!hb]
    \centering
    \includegraphics[width = 0.9\textwidth]{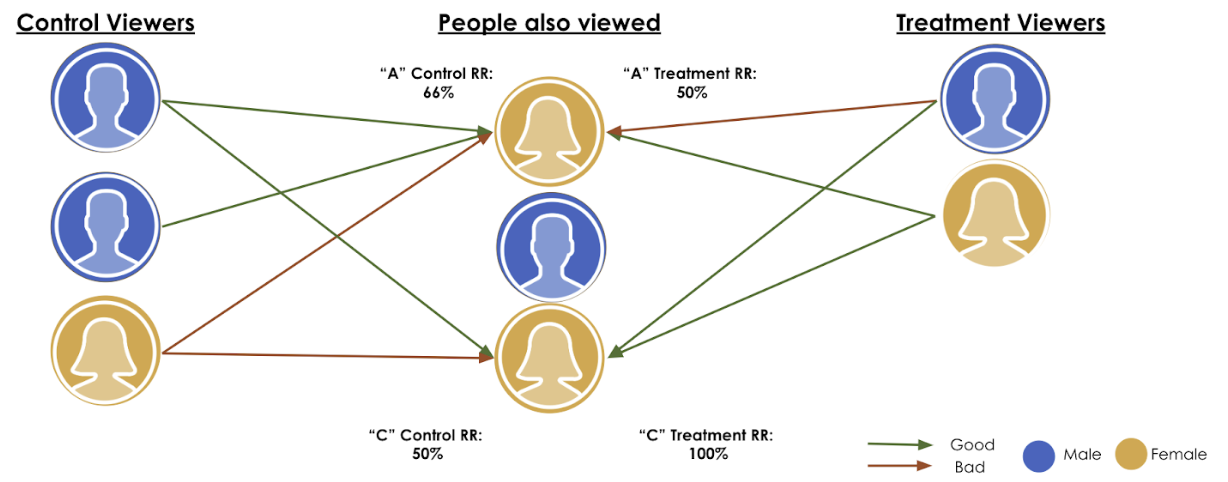}
    \caption{We look at all edges from control to females and treatment to female and check if those edges are ``good" at a similar rate. ``Good" edges are formed when the recommended member accepts the invite, replies to a message and doesn't report the viewer. ``Bad" edges are formed on the contrary. RR stands for the response rate.}
    \label{fig:edge-level-analysis}
\end{figure*}

Figure \ref{fig:edge-level-analysis} shows a toy example to highlight the analysis. 

We compute the edge-level response rates (RR) for the female members “A” and “C.” Green edges indicate positive responses (e.g. a connection invite was sent in the arrow direction and the invite was accepted) while red edges indicate negative responses (e.g. ignored/rejected invite). Our aim is to detect if the positive response rate to treatment viewers is lower, which may indicate that females are receiving lower-quality interactions on the platform.

The full edge-based analysis results are highlighted in the Table \ref{tab:edge}.

\begin{table*}[!hb]\centering
\begin{tabular}{lrrrrr}\toprule
\textbf{Edge Type} &\textbf{Metric} &\textbf{ControlRate} &\textbf{TreatmentRate} &\textbf{RelativeDiff} \\\midrule
All Edges &accept &0.211385 &0.244152 &0.155011 \\
Edges to Female &accept &0.20275 &0.234024 &0.154251 \\
Edges to Male &accept &0.218781 &0.250646 &0.145647 \\
All Edges &reply &0.313753 &0.305175 &-0.027337 \\
Edges to Female &reply &0.312765 &0.306352 &-0.020507 \\
Edges to Male &reply &0.310661 &0.300983 &-0.031154 \\
All Edges &report &7.00e-5 &7.60e-5 &0.077257 \\
Edges to Female &report &4.70e-5 &4.60e-5 &-0.017529 \\
Edges to Male &report &8.30e-5 &9.20e-5 &0.115913 \\
\bottomrule
\end{tabular}
\caption{Edge-level response rates across metrics comparing the base model with the BMT model.}\label{tab:edge}
\end{table*}

\section{Self-ID}
\begin{figure*}[!ht]
     \centering
     \begin{subfigure}[b]{0.3\textwidth}
         \centering
         \includegraphics[width=\textwidth]{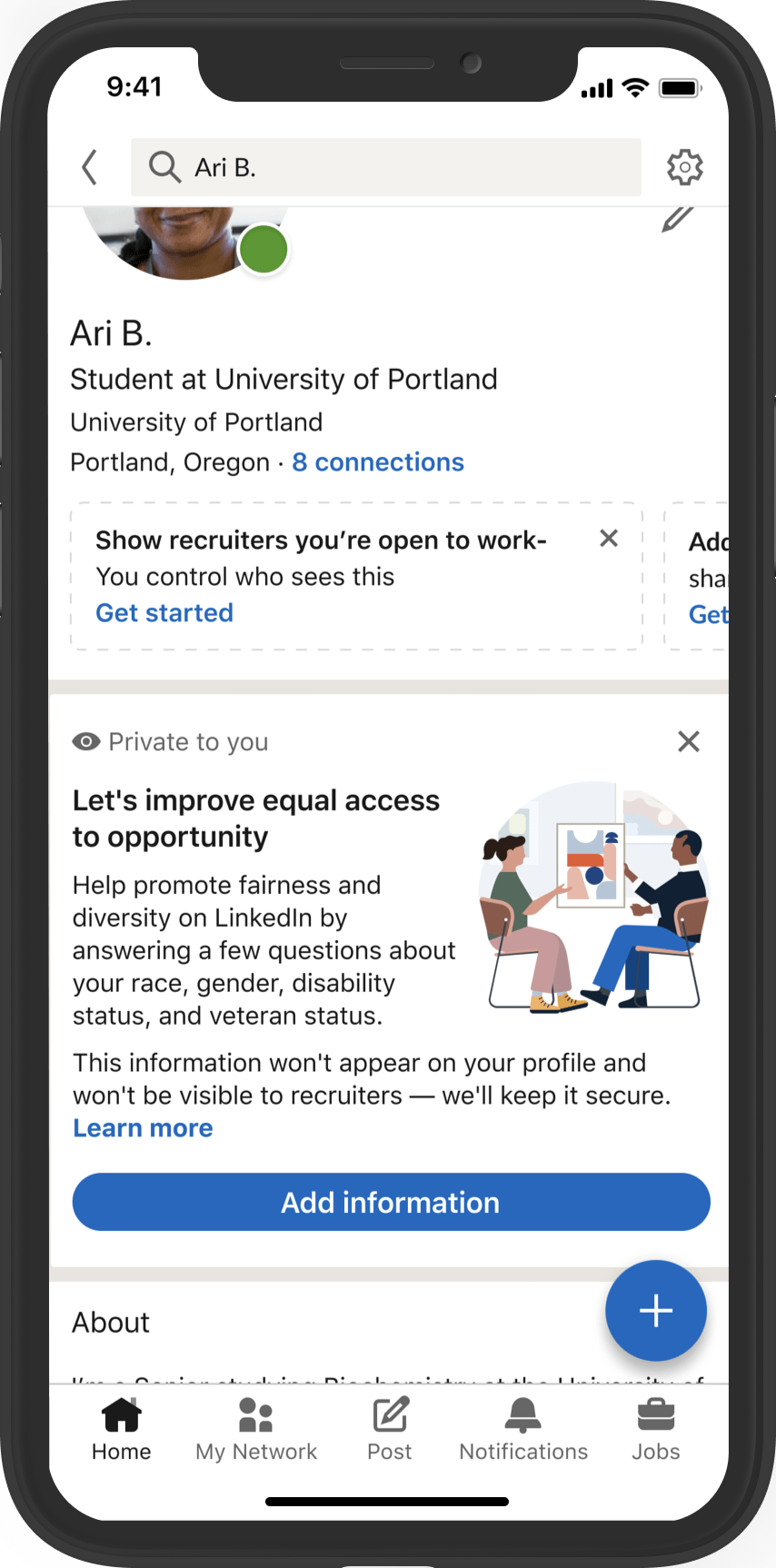}
     \end{subfigure}
     \hspace{2cm}
     \begin{subfigure}[b]{0.43\textwidth}
         \centering
         \includegraphics[width=\textwidth]{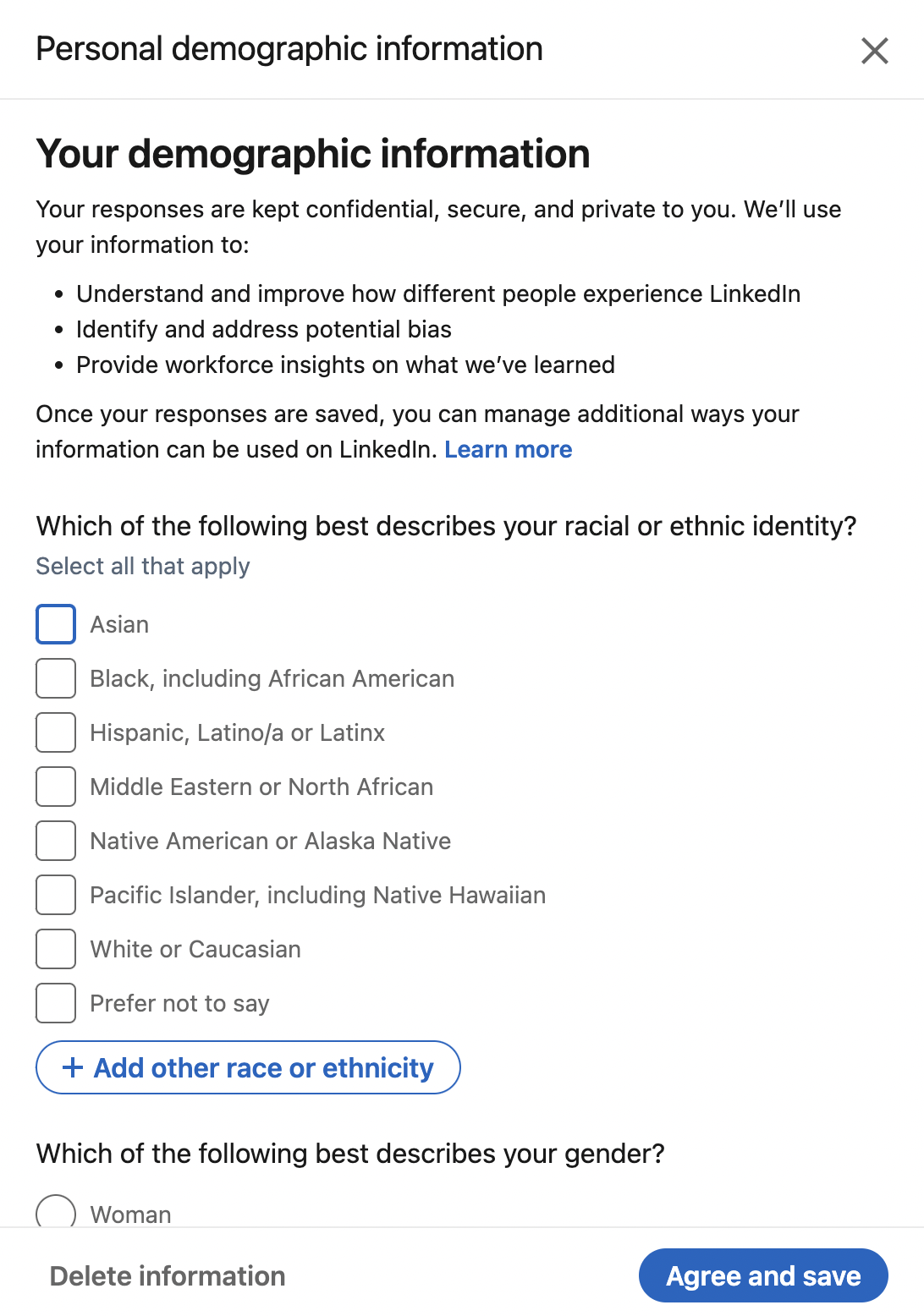}
     \end{subfigure}
        \caption{Self-ID form (partially visible) which is accessible through the profile update feature on LinkedIn}
        \label{fig:selfid}
\end{figure*}

Figure \ref{fig:selfid} shows one of the entry points into the form as well as some of the initial questions in the form. For more details on self-id please see \cite{selfid1, selfid2, selfid3}.

\section{People Also Viewed}
This People Also Viewed product recommends additional profiles you might be interested in based on the current profile you are viewing and is an example of the scarce resource allocation paradigm when considering the recommended members. The product is the second largest source of traffic for profile views, therefore any negative bias (failing to surface qualified recommendations) could adversely affect members’ abilities to grow their network.

\begin{figure*}[!ht]
    \centering
    \includegraphics[width = \textwidth]{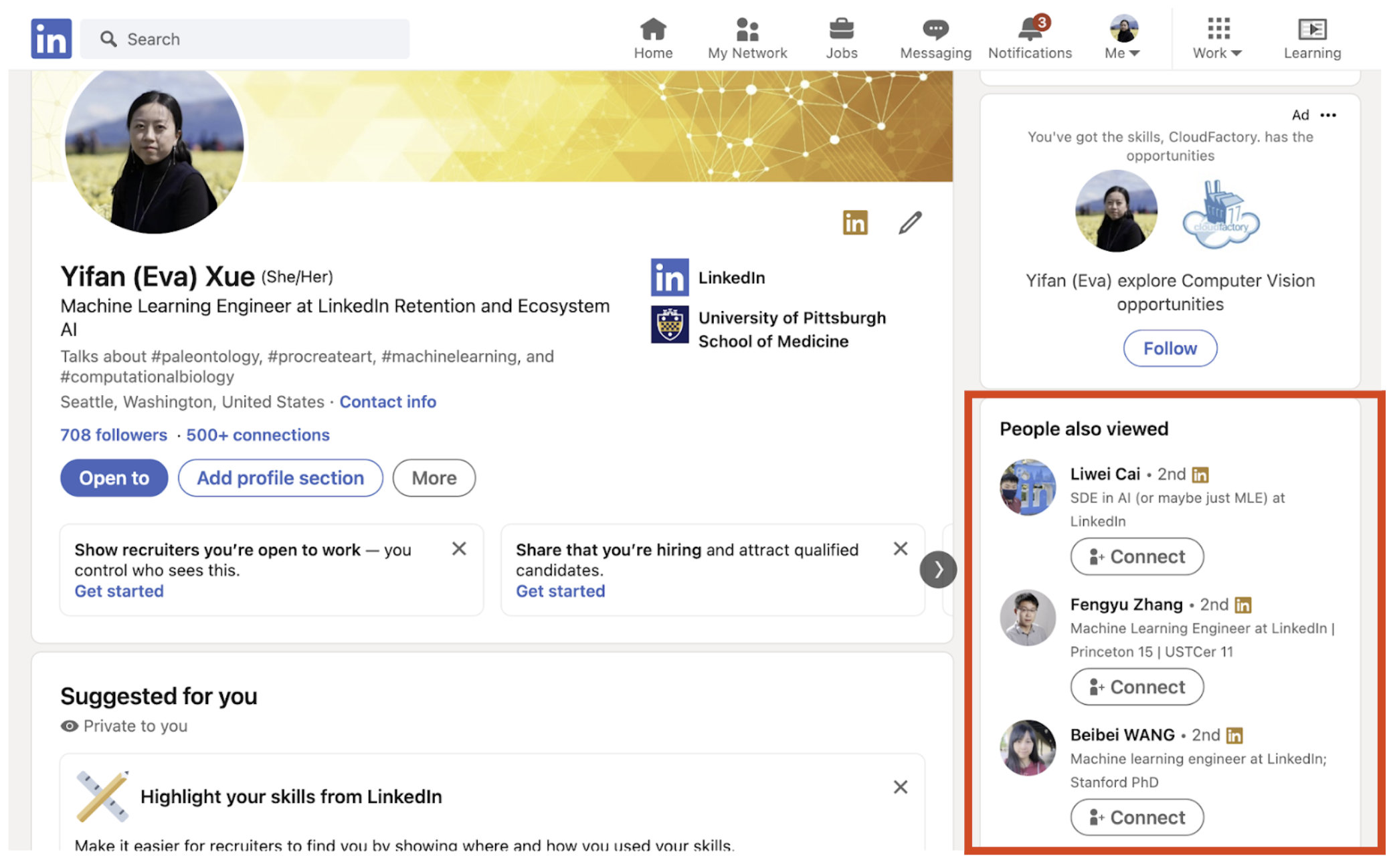}
    \caption{In this example, Yifan Xue is the source member, the recommended members in the highlighted “People also viewed” card are dest members, and whoever is viewing the profile is the viewer.}
    \label{fig:pav_full}
\end{figure*}

The list of recommendations appear when a viewer is looking at a particular member's (source) profile. See the Figure \ref{fig:pav_full} for the details.

\end{document}